\documentclass[twocolumn,superscriptaddress,secnumarabic,amssymb, nobibnotes, aps, bm,pre,floatfix,longbibliography]{revtex4-2} 
\usepackage[english]{babel}
\usepackage[latin9]{inputenc}
\usepackage{graphicx}
\usepackage{multirow}
\usepackage[usenames,dvipsnames]{xcolor}
\usepackage{bm}
\usepackage{mathtools}
\usepackage{amsmath}
\usepackage{rotating}
\definecolor{oceanboatblue}{rgb}{0.0, 0.47, 0.75}
\definecolor{orange}{rgb}{1,0.5,0}
\definecolor{goodgreen}{rgb}{0.1,0.5,0}
\definecolor{goodred}{rgb}{0.7,0,0}
\usepackage[colorlinks,urlcolor=goodgreen,citecolor=blue,linkcolor=goodred]{hyperref}
\usepackage{cleveref}
\usepackage{lineno}
\setpagewiselinenumbers
\modulolinenumbers[5]
\usepackage{tikz}
\usepackage{tocvsec2}
\usepackage{scrextend}
\usepackage{braket}
\makeatletter
\newcommand{\ii}{\text{i}}
\newcommand{\ee}{\text{e}}
\usepackage[normalem]{ulem}

\makeatother

\usepackage[normalem]{ulem}

\begin{document}

\title{Implementation and characterization of the dice lattice in the electron quantum simulator}
\author{C. Tassi}
\email{camillo.tassi@gmail.com}
\affiliation{Donostia International Physics Center (DIPC), 20018 Donostia--San Sebasti\'an, Spain}
\affiliation{Dipartimento di Fisica dell'Universit\`a di Pisa, Largo Bruno Pontecorvo 3, I-56127 Pisa, Italy}

\author{D. Bercioux}
\email{dario.bercioux@dipc.org}
\affiliation{Donostia International Physics Center (DIPC), 20018 Donostia--San Sebasti\'an, Spain}
\affiliation{IKERBASQUE, Basque Foundation for Science, Plaza Euskadi 5
48009 Bilbao, Spain}

\begin{abstract}
Materials featuring touching points, localized states, and flat bands are of great interest in condensed matter and artificial systems due to their implications in topology, quantum geometry, superconductivity, and interactions. In this theoretical study, we propose the experimental realization of the dice lattice with adjustable parameters by arranging carbon monoxide molecules on a two-dimensional electron system at a (111) copper surface. First, we develop a theoretical framework to obtain the spectral properties within a nearly free electron approximation and then compare them with tight-binding calculations. Our investigation reveals that the high mobility of Shockley state electrons enables an accurate theoretical description of the artificial lattice using a next-nearest-neighbor tight-binding model, resulting in the emergence of a touching point, a quasi-flat band, and localized lattice site behavior in the local density of states. Additionally, we present theoretical results for a long-wavelength low-energy model that accounts for next-nearest-neighbor hopping terms. Furthermore, we theoretically examine the model's behavior under an external magnetic field by employing Peierl's substitution, a commonly used technique in theoretical physics to incorporate magnetic fields into lattice models. Our theoretical findings suggest that, owing to the exceptional electron mobility, the highly degenerate eigenenergy associated with the Aharonov-Bohm caging mechanism may not manifest in the proposed experiment.
\end{abstract}
\date{\today}
\maketitle
\section{Introduction}

In the past few years, there has been a growing interest in studying lattice systems characterized by spectra containing one or more quasi-flat bands~\cite{Toermae_2022}. This interest stems from the discovery of superconductivity and other correlated phases in layered systems~\cite{Andrei_2020,Kennes_2021}. The prevailing understanding suggests that correlated phenomena are observable in lattice systems with quasi-flat bands, primarily due to the careful comparison of energy scales. Specifically, correlation effects become visible as electronic kinetic energy diminishes to the same order of magnitude as, or even below, the electron interaction energy, thus highlighting an intricate interplay of energy scales at play.

However, the interest of the condensed matter community in the properties of lattice systems with flat bands roots back to the past 30 years, when the effects of frustration and destructive interference were investigated in the kagome~\cite{Baxter_1970}, the dice~\cite{sutherland1986localization}, and the Lieb lattice~\cite{Lieb}. All three of these lattice systems are characterized by a perfect flat band that is a finite energy for the case of the kagome lattice and at zero energy for the latter two lattices. The bulk wave function of these flat bands can be expressed as a linear combination of the so-called \emph{compact localized states} (CLSs)~\cite{Aoki_1996,Bergman_2008,Rhim_2019}. These states, corresponding to wave functions with support only on a finite number of lattice sites, can be regarded as a third possible type of basis alongside the Bloch and Wannier bases. However, it's important to note that CLSs do not necessarily constitute a complete and orthogonal set~\cite{Rhim_2019}. The CLSs associated with a flat band exhibit high degeneracy and possess intriguing quantum geometry properties~\cite{Graf_2021}. 
Signatures of CLSs have been observed in different systems presenting flat bands, as the photonic diamond chain~\cite{Mukherjee_2018} and photonic Lieb lattice~\cite{Mukherjee_2015}, and in the electronic realm in the quasi-1D diamond-necklace chain~\cite{Kempkes_2023}.

Photonic and electronic platforms represent ideal quantum simulators of single particle physics~\cite{Aspuru-Guzik_2012,Khajetoorians_2019}. 
The concept of quantum simulators can be traced back to a suggestion by Richard Feynman~\cite{Feynman_1982}. He proposed using a controllable quantum system to simulate another quantum system with unknown properties. Since then, many quantum simulation platforms have been developed~\cite{Altman_2021}. In this article, we will mainly consider a specific condensed-matter quantum simulator, also known as the \emph{electron} quantum simulator~\cite{Khajetoorians_2019}. In this type of platform, the free electrons of a metallic surface of a noble metal are unambiguously constrained via atoms or molecules to move along a potential profile meticulously designed to resemble one of the specific lattices under investigation, the properties of which are yet to be determined. Within this technique, it has been possible to explore the physics of graphene~\cite{gomes2012designer, Wang_2014}, aperiodic structures as fractals~\cite{Kempkes_2019_b}, and the Pensore tiling~\cite{Collins2017}, in addition to many other different lattice, as summarized in Refs.~\cite{Khajetoorians_2019,Freeney_2022,Piquero-Zulaica_2022}.

%
%
\begin{figure*}
\centering
\includegraphics[width=0.88\textwidth]{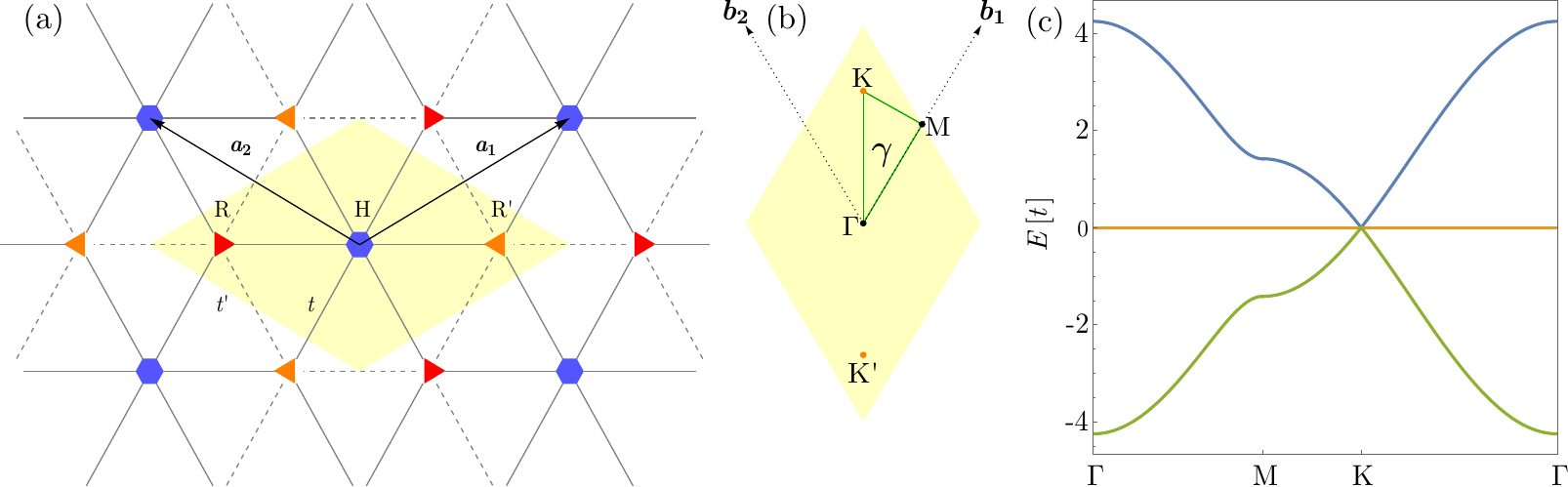}
\caption{\label{fig_1}(a) The next-nearest-neighboring dice lattice is a hexagonal-symmetry lattice. The unit cell (denoted with a yellow rhombus) contains three sites, indicated with H (blue hexagon), R (red triangle), and R' (orange triangle). The vectors arrows $\bm{a}_1$ and $\bm{a}_2$ are the lattice vectors. The solid and dashed black lines denote the nearest-neighboring hopping term $t$ and the next-nearest-neighboring hopping term $t'$, respectively.  (b) The dice lattice first Brillouin zone, $\bm{b}_1$, and $\bm{b}_2$ are the reciprocal vectors indicated; these are defined via the relation $\bm{a}_i\cdot\bm{b}_j=2\pi\delta_{i,j}$. The solid green line is the high symmetry path $\gamma$. (c) The band structure along the high symmetry path. The energies are measured in units of $t$ and the lengths in units of $a$.}
\end{figure*}
%
Of particular interest for this manuscript are the proposal and the realization of the Lieb lattice in the electron quantum simulator~\cite{qiu2016designing,slot2017experimental,Drost_2017,Bercioux_2017_b}. The reason boils down to the spectral properties of the Lieb lattice; this is a face-centered square lattice characterized by three energy bands. In the simplest next-neighbor tight-binding approximation, the energy spectrum of the Lieb lattice is energy symmetric, with one of the three bands completely flat and at zero energy. The three bands share an accidental three-fold degeneracy at the M point of the first Brillouin zone (BZ). For this momentum, it is possible to present a long-wavelength low-energy approximation for the Hamiltonian of the Lieb lattice, resulting in an effective pseudo-spin one Hamiltonian~\cite{Goldman_2011}. Additionally, the flat band of the Lieb lattice can be interpreted in terms of CLSs~\cite{Aoki_1996,Rhim_2019}.

Here, we will investigate and propose a possible experimental implementation of another lattice with a flat band at zero energy: the dice lattice~\cite{sutherland1986localization}. We present a sketch of the dice lattice in \textbf{Figure}~\ref{fig_1}(a). It is a triangular lattice with a base of three sites in the unit cell; as in the case of the Lieb lattice, the connectivity of the three sites is not homogeneous. Specifically, we have one site, dubbed H with connectivity 6, and two inequivalent sites, named R and R' with connectivity 3. Similar to the Lieb lattice, the energy spectrum of the dice lattice consists of three bands, two of which are energy symmetric, and the third one is completely flat and at zero energy. The three bands have an accidental degeneracy at the K point of the BZ. In this point, it is possible to describe the long-wavelength low-energy expansion of the system Hamiltonian in terms of an integer pseudo-spin theory~\cite{Bercioux_2009}. Some previous work proposed the experimental implementation the dice lattice in the framework of cold atoms in optical lattices~\cite{Bercioux_2009,Bercioux_2011}.

A peculiar property of the dice lattice compared to the Lieb one is the evolution of the energy spectrum under the application of a perpendicular magnetic field. In the absence of disorder, the dice lattice exhibits extreme wavefunction localization, leading to a peculiar energy spectrum characterized by three perfectly flat bands: two with finite and opposite energies and a third at zero energy. This localization phenomenon, known as Aharonov-Bohm caging \cite{vidal1998aharonov, Vidal_2000, vidal2001disorder}, arises from the interplay between lattice geometry and the applied magnetic field, contrasting with the ordinary Anderson localization occurring in a disordered system. A few experiments have shown the signature of the Aharonov-Bohm caging in the dice lattice~\cite{Abilio_1999,Naud_2001,Li_2022,Martinez_2023}.

In this article, we propose a realization of the dice lattice in the electron quantum simulator. Specifically, we will consider the case of decorating with CO molecules the surface (111) of  Cu~\cite{gomes2012designer,Khajetoorians_2019}. We propose in total three different realizations differing for the amount of CO molecules implemented and the area of the unit cell. We evaluate the spectral properties of these configurations and show how they map into the simplest next-nearest-neighboring tight-binding approximation. The advantage of implementing the dice lattice compared to the Lieb lattice roots is down to the compatibility between the space group of the dice lattice with one of the Cu(111) substrates. It is important to note that the set of parameters we used for our implementation are in agreement with the experimental parameters~\cite{gomes2012designer,slot2017experimental,Kempkes_2019,Kempkes_2019_b,Freeney_2022}.

The paper is organized as follows. In \textbf{Section}~\ref{dice_TB}, we revise the properties of the dice lattice spectrum in the next-nearest-neighboring tight-binding approximation. In \textbf{Section}~\ref{proposal}, we present our three proposals for the experimental implementation and find the corresponding parameters to map into next-nearest-neighboring tight-binding models. In \textbf{Section}~\ref{lwa}, we revise the long-wavelength low-energy approximation for the dice lattice Hamiltonian expanded around the K point of the BZ, including terms describing the next-nearest-neighboring hopping terms. Finally, in \textbf{Sections}~\ref{butterfly_tb}, we revise the properties of the dice lattice in a perpendicular magnetic field in the tight-binding approximation. In \textbf{Sections}~\ref{conclusions}, we summarize the work and give some possible outlook. In \textbf{Appendix}~\ref{app_I}, we show what is happening when changing the system parameters compared to the standard one used in the experiments~\cite{gomes2012designer,slot2017experimental,Kempkes_2019,Kempkes_2019_b,Freeney_2022}.

\section{Results}\label{Sec_results}
In the following, we describe the general properties of the dice lattice within the simplest tight-binding model without a magnetic field. Successively, we present a possible theoretical implementation with an electron quantum simulator platform based on CO molecules on Cu(111). The numerical implementation is based on the nearly-free electron model, in which a muffin-tin potential models the CO molecules. We compare the results with a next-nearest-neighbor tight-binding model. Finally, we present the results for the spectrum when a perpendicular magnetic field is applied.

\subsection{Tight-binding model}\label{dice_TB}

The dice lattice is a regular tessellation of the Euclidean plane with regular $\pi/3$ rhombi, resulting in a bipartite structure with hexagonal symmetry. Each unit cell contains three sites; in the following, we name them a hub, indicated with the letter H, and two rims, indicated with R and R', respectively~\cite{vidal1998aharonov}. We present a sketch of the lattice in \textbf{Figure}~\ref{fig_1}(a). The following translational vectors characterize the unit cell:
%
%
\begin{align}
    \bm{a}_1= \frac{a}{2}\left(3,\sqrt{3}\right), \quad \bm{a}_2= \frac{a}{2}\left(-3,\sqrt{3}\right),
\end{align}
%
%
where $a$ is the distance between the lattice sites. In the following, we consider a rhombic unit cell for simplicity, while the Wigner-Seitz cell is hexagonal. We describe the dice lattice with a simple $s$-orbital tight-binding model, including next-neighbor (solid lines in \textbf{Figure}~\ref{fig_1}(a)) and next-nearest-neighboring (dashed lines in \textbf{Figure}~\ref{fig_1}(a)) hopping terms, $t$ and $t'$, respectively. 
Neglecting the next-nearest-neighboring hopping terms $t'$, the hub sites are sixfold connected, while the rim sites are threefold connected. Thus, including $t'$ terms will destroy this property, and all the lattice sites will have the same connectivity. This is important since the flat band originates in unequal connectivity among the sites.

The Bloch Hamiltonian that includes both hopping terms reads:
%
%
\begin{align}\label{BlochHam}
\hat{h}(\bm{k})=
  \begin{pmatrix}
   0 &&
   f_\text{RH}(\bm{k})\, t &&
   f_\text{RR'}(\bm{k})\, t' \\
    f_\text{RH}^\ast(\bm{k})\, t &&
   -\Delta &&
   f_\text{HR'}(\bm{k})\, t\\
    f_\text{RR'}^\ast(\bm{k})\, t'&&
   f_\text{HR'}^\ast(\bm{k})\, t&&
   0
  \end{pmatrix}
 \end{align}
%
%
expressed in the following base $\Psi=(\psi_\text{R},\psi_\text{H},\psi_\text{R'})^\text{T}$. Here, we have introduced the following auxiliaries functions: 
%
%
\begin{subequations}
\begin{align}  
  f_\text{RH}(\bm{k})&= 1+\ee^{-\ii\, \bm{k}\cdot\bm{a}_2} +\ee^{-\ii\, (\bm{k}\cdot\bm{a}_1+\bm{k}\cdot\bm{a}_2)},\\
  f_\text{HR'}(\bm{k})&=1+\ee^{-\ii\, \bm{k}\cdot\bm{a}_1} +\ee^{-\ii\, (\bm{k}\cdot\bm{a}_1+\bm{k}\cdot\bm{a}_2)},\\
  f_\text{RR'}(\bm{k})&=1+\ee^{-\ii\, \bm{k}\cdot\bm{a}_1 } +\ee^{\ii\, \bm{k}\cdot\bm{a}_2},
\end{align}
\end{subequations}
%
%
where $\bm{k}$ is the crystal momentum vector defined within the BZ, see \textbf{Figure}~\ref{fig_1}(b). In Eq.~\eqref{BlochHam}, we have introduced an energy term $\Delta$ describing the difference between the hub and rim on-site energy. It is not possible to find a compact expression for the energy bands; however, for $t'=0$, the energy spectrum is given by
%
%
\begin{subequations}\label{spectrum_tp_0}
    \begin{align}
    \mathcal{E}_\text{c}&=0,\\
    \mathcal{E}_\alpha(\bm{k})&=\alpha t\sqrt{6 + \left(\frac{\Delta}{4 t}\right)^2+4 \Lambda(\bm{k})}-\frac{\Delta}{2},\label{spectrum_dispersive}
\end{align}
\end{subequations}
%
%
%
%
\begin{figure}
    \centering
    \includegraphics[width=0.9\columnwidth]{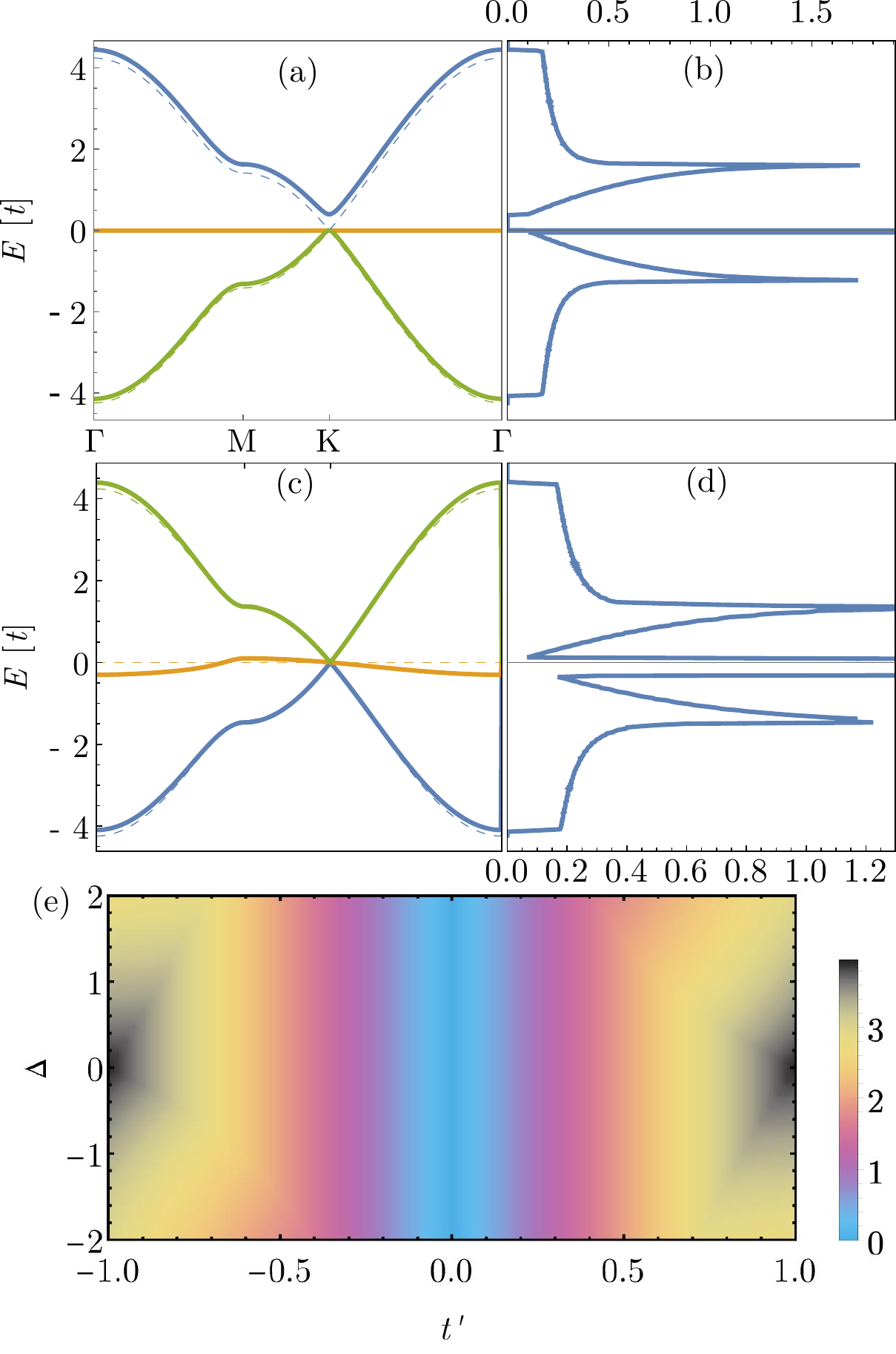}
    \caption{ \label{fig_2} (a) The spectrum of the dice lattice as a function of the momentum $k$ along a high symmetry path for $\Delta=-0.4t$ and $t'=0$, (b) Density of the states for the same parameters as in panel (a). (c) Dice lattice spectrum as a function of the momentum $k$ along a high symmetry path for $\Delta=0$ and $t'=0.1$, (d) Density of the states for the same parameters as in panel (c). (e) Bandwidth of the central band $\mathcal{E}_\text{c}$ as a function of $\Delta$ and $t'$, these parameters are measured in units of $t$. In panels (a) and (c), the dashed lines correspond to the spectrum for $\Delta=t'=0$,  as in Fig.~\ref{fig_1}(c).}
 \end{figure}
%
%
where we have introduced the function 
%
%
\begin{align*}
\Lambda(\bm{k})&=\cos[\bm{k}\cdot \bm{a}_1]+\cos[\bm{k}\cdot \bm{a}_2]+\cos[\bm{k}\cdot(\bm{a}_1+\bm{a}_2)] \\
& = 2\cos\left(\frac{3 k_x a}{2}\right)\cos\left(\frac{\sqrt{3}k_y a}{2}\right)+\cos(\sqrt{3}k_y a)\,.
\end{align*}
%
%
%
%
\begin{figure*}
\centering
{\includegraphics[width= .99\textwidth]{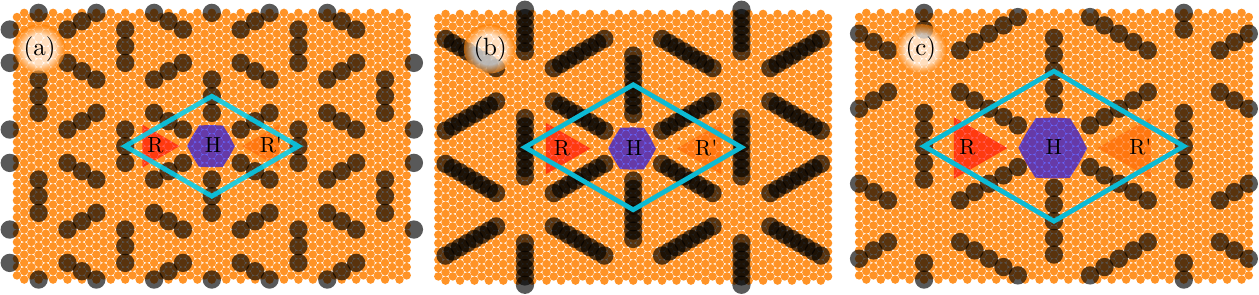}}
\caption{\label{Fig_3} Different implementations of the dice lattice in order of unit cell area. The orange disks represent the copper atoms of the Cu(111) triangular lattice surface. Black disks represent the CO molecules on top of the copper atoms. The blue solid line delimits the unit cell and contains the hub site H and the two rim sites R and R'. In (a), the unit cell contains nine CO molecules, and the unit cell area is about $8\, \text{nm}^2$  | CI.  In (b), the unit cell contains twenty-four molecules, and the unit cell area is about $13\, \text{nm}^2$ | CII.  In (c),  the unit cell contains fifteen molecules, and the unit cell area is about $18\, \text{nm}^2$  | CIII.}
\end{figure*}
%
%
We start presenting the spectral properties of the dice lattice in the simple case of $t'=\Delta=0$. In this limit, of the three bands composing the spectrum, two are dispersive $\mathcal{E}_\pm$, and one is completely flat at zero energy $\mathcal{E}_\text{c}$ | \textbf{Figures}~\ref{fig_1}(c). The two dispersive bands are identical to the one of the honeycomb lattice~\cite{CastroNeto_2009}. The flat band $\mathcal{E}_\text{c}$ arises because of the bipartite nature of the dice lattice~\cite{Vidal_2000,Bergman_2008} and can be interpreted in terms of CLSs as well~\cite{Zhang_2020}. The three bands are degenerate in the K and K$'$ points of the BZ. This threefold degeneracy is accidental since, in two dimensions, there are no irreducible representations of order three~\cite{Bradlyn_2016,Bradlyn_2017}. Performing a low-energy long-wavelength approximation of the energy spectrum around this K or K$'$, it is possible to obtain an effective pseudo-spin one Hamiltonian model for the system~\cite{Bercioux_2009,Bercioux_2011,Goldman_2011,Urban_2011}, see \textbf{Section}~\ref{lwa} for additional details.

By inspecting Eq.~\eqref{spectrum_dispersive}, we notice that the onsite energy $\Delta$ shifts the two dispersive bands and partially removes the degeneracy at the K(K$'$)-points by opening a gap between a dispersive band and the flat one | \textbf{Figure}~\ref{fig_2}(a). Changing the sign of $\Delta$, the energy gap opens between the flat band and the other dispersive one. Thus, we conclude that this term breaks the three-fold degeneracy at the K(K$'$)-points. Furthermore, a finite value of the next-nearest-neighboring hopping $t'$ induces a finite dispersion in the flat band | \textbf{Figure}~\ref{fig_2}(b), but it does not open an energy gap at the three-fold degeneracy point in $K$.  The energy symmetry is broken when $\Delta$ and $t'$ are finite | see \textbf{Figure}~\ref{fig_1}(d).
We note on passing that the effects of $t'$ are similar to the effects of evaluating the energy spectrum by considering an overlap integral~\cite{Goerbig_2011}.

In \textbf{Figures}~\ref{fig_2}(b) and~\ref{fig_2}(d), we present the density of states (DOS) corresponding to the spectra depicted in \textbf{Figures}~\ref{fig_2}(a) and~\ref{fig_2}(c), respectively. We define the DOS as
%
%
\begin{equation}\label{eq_DOS}
    \rho(\epsilon)=\sum_n \int_{\text{BZ}} \frac{d\bm{k}}{2 \pi^2}\, \delta(\epsilon-\epsilon_{n \bm{k}}),
\end{equation}
%
%
where $\epsilon_{n \bm{k}}$ is the energy of the band $n$ for the two-dimensional momentum~$\bm{k}$ within the BZ.
Notably, it features characteristic Van Hove singularities, indicative of the saddle points at the M point in the energy spectrum, see \textbf{Figure}~\ref{fig_2}(b), whereas, when $t'\neq0$, it features a double peak one close to the M point and another one close in energy, see \textbf{Figure}~\ref{fig_2}(d). Additionally, in both cases, it exhibits a distinct peak associated with the flat or quasi-flat band $\mathcal{E}_\text{c}$.

Finally, we present in \textbf{Figure}~\ref{fig_2}(e) the bandwidth evolution of the $\mathcal{E}_\text{c}$ band as a function of $\Delta$, and $t'$. We have defined the bandwidth as difference between the maximum and the minumum value of $\mathcal{E}_\text{c}(\bm{k})$, with $\bm{k}$ in the BZ. Here, we can observe that the bandwidth grows linearly with the next-nearest-neighboring hopping $t'$ up to values of this of the order of $t'\sim 0.7 t$, then it starts to change slope, especially for large values of the onsite energy $\Delta$. Additionally, we note that for $t'<t/2$, the bandwidth is independent of the onsite energy $\Delta$.

\subsection{Implementation with CO molecules on Cu(111)}\label{proposal}

In this section, we explore diverse implementations of the dice lattice within the electron quantum simulator. These are realized by decorating the Cu(111) surface with CO molecules~\cite{Khajetoorians_2019,Freeney_2022,Piquero-Zulaica_2022}. The Cu(111) surface is characterized by a triangular lattice structure featuring an interatomic distance of $a_\text{Cu}=0.256$~nm. Specifically, we introduce three distinct configurations for the dice lattice, as illustrated in \textbf{Figure}~\ref{Fig_3}, each distinguished by the number of CO molecules utilized to construct the unit cell. In all proposed configurations, CO molecules play a crucial role in designing the unit cell's lattice sites with varying connectivity and in tuning the hopping probabilities between the rim sites. This tunneling is possible since electrons can tunnel beneath the CO potential barriers, enabling a nonzero next-nearest-neighbor hopping~$t'$~\cite{Freeney_2022}. It is noteworthy that the sizes of the three sites within the lattice are not uniform, leading to different onsite energies for the hub and the rim sites.

In the following, we summarize the key steps for obtaining the spectral properties for these three dice lattice implementations with CO molecules. In general, we want to find the spectral properties of the following Hamiltonian
%
%
\begin{equation}
    H=\frac{\bm{p}^2}{2m^*}+V_\text{CO}(\bm{R}),
\end{equation}
%
%
where $m^*$ is the effective mass of the free electrons on Cu(111)~\cite{Freeney_2022}, the periodic potential $V_\text{CO}(\bm{R})$ is obtained by considering the set of all CO molecules. Without loss of generality, we can assume that each CO molecule at the position $\bm{r}_0$ is a cylindrical repulsive potential for the Cu(111) surface electrons~\cite{gomes2012designer}:
%
%
\begin{equation}\label{cylindrical_pot}
V(\bm{r}-\bm{r}_0) = V_0 \begin{cases}
1 & |\bm{r}- \bm{r}_0|
\leqslant \frac{D}{2} \\
0 & |\bm{r}- \bm{r}_0|> \frac{D}{2}
\end{cases}.
\end{equation} 
%
%
The parameters of this potential are chosen in agreement with the experiments as $V_0=0.9\, \text{eV}$ and a cylinder diameter of $D=0.6\, \text{nm}$~\cite{gomes2012designer,slot2017experimental,Freeney_2022}. In \textbf{Appendix}~\ref{app_I}, we show what happens by changing the shape of the potential describing the CO molecules and the behavior by changing the system parameters away from the optimal ones. The complete set of CO molecules can be described by a periodic potential $V_\text{CO}(\bm{R})$ that can be expanded in terms of Fourier components  as
%
%
\begin{equation}\label{periodic_pot}
V_\text{CO}(\bm{R})=\sum_{\bm{G}} V_{\bm{G}}\, \ee^{\ii \bm{G}\cdot\bm{R}},
\end{equation}
%
%
here $\bm{G}$ are vectors of the reciprocal lattice.
Because of the shape of each cylindrical potential in Eq.~\eqref{cylindrical_pot}, the Fourier components $V_{\bm{G}}$ are
%
%
\begin{subequations}
\begin{align}\label{Fourier_cyl}
V_{\bm{G}}& = \int_\text{UC} d\bm{r}\ V(\bm{r}-\bm{r}_0)\text{e}^{-\text{i} \bm{G}\cdot \bm{r}} \nonumber\\ 
& = (2\pi r_0 V_0) \frac{\mathcal{J}_1(|\bm{G}|D)}{|\bm{G}|}\,,
\end{align}
\end{subequations} 
%
%
where $\mathcal{J}_1(x)$ are the Bessel functions of the first order, and the integral is performed over the unit cell (UC). Summing up over all the Fourier components $V_{\bm{G}}$, we can rewrite Eq.~\eqref{periodic_pot} as:
%
%
\begin{align}\label{exact_potential}
V_\text{CO}(\bm{R}) & = \frac{2\pi r_0 V_0}{A} \sum_{\bm{G}}\frac{\mathcal{J}_1(|\bm{G}|r_0)}{|\bm{G}|} \text{e}^{\text{i} \bm{G}\cdot\bm{R}} \\
& = \frac{2\pi r_0 V_0}{A}\!\!\!\! \sum_{\{\ell,m\}\in\mathbb{Z}}\!\!\frac{\mathcal{J}_1(|\ell\bm{b}_1+m\bm{b}_2|D)}{|\ell\bm{b}_1+m\bm{b}_2|} \text{e}^{\text{i} (\ell\bm{b}_1+m\bm{b}_2)\cdot\bm{R}}. \nonumber
\end{align}
%
%
%
%
\begin{figure*}
\centering
{\includegraphics[width= .99\textwidth]{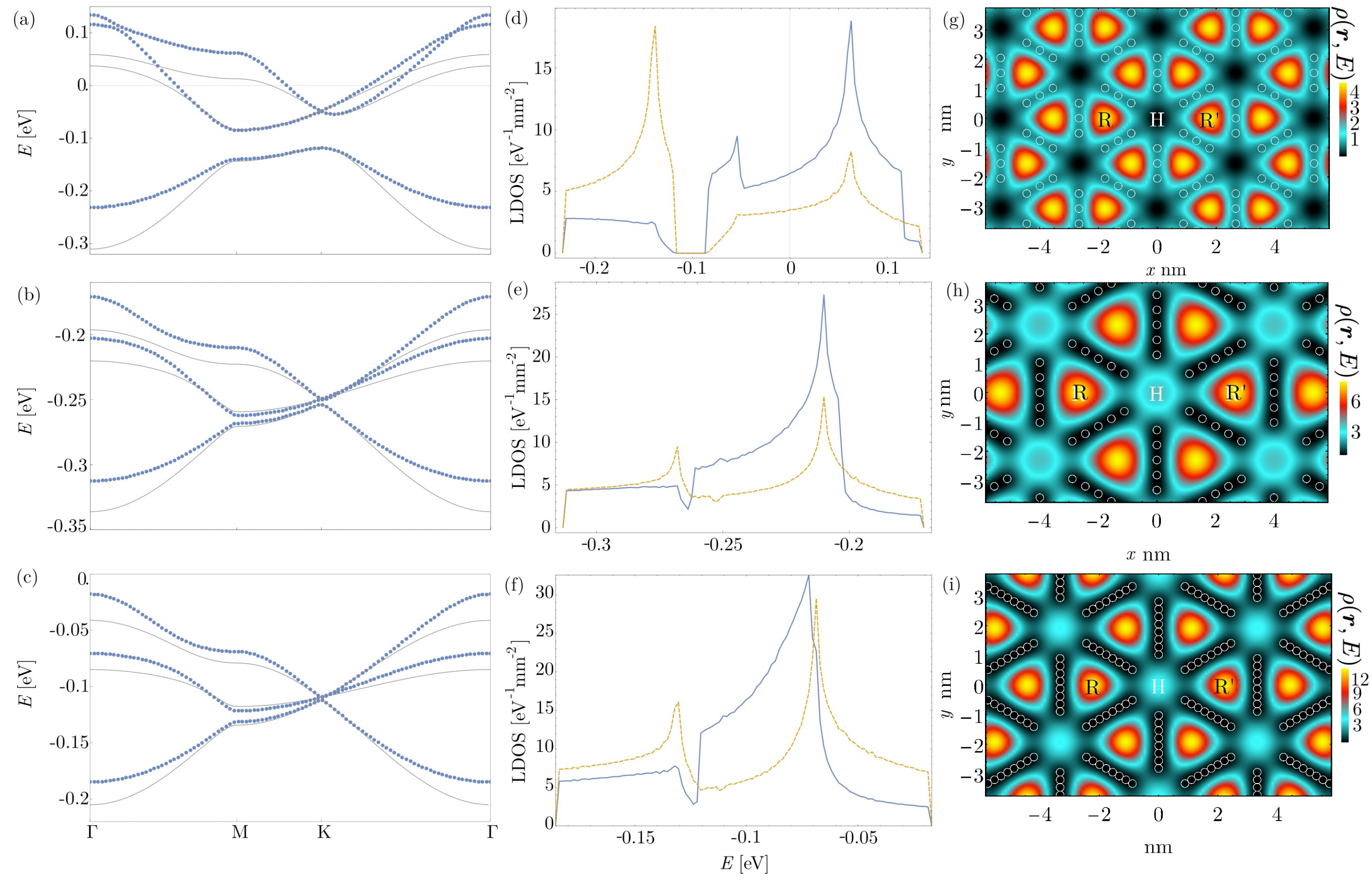}}
\caption{\label{Fig_4} In the left column [Panels (a), (d), and (g)], energy spectra for the three configurations in \textbf{Figure}~\ref{Fig_3} (dotted lines) and energy spectra obtained by the eigenvalues of Eq.~\eqref{BlochHam} with the parameters presented in \textbf{Table}~\ref{Tab_1}.
In the central column [Panels (c), (f), and (i)], the LDOS calculated at the center of the rim (dashed line) and hub (dot-dashed line) sites as a function of the energy (the Fermi energy is set at $E=0$). In the right column [Panels (b), (e), and (h)], the LDOS density plot for energies around the middle bands. The circles indicate the CO molecules' position.  The three rows refer to the configurations in \textbf{Figure}~\ref{Fig_3}. }
\end{figure*}
%
%
At this point, assuming that the system wave function is also periodic~\cite{Ashcroft_1976}, we can transform the Schr\"odinger equation in a set of coupled linear equations
%
%
\begin{align}\label{sys_equations}
\left[\frac{\hbar^2 \left|\bm{q}-\bm{G}\right|^2}{2m}-\mathcal{E}_n(\bm{q})\right]&c^n_{\bm{q-G}} \\
&\hspace{-1cm}+ \sum_{\bm{G'}}  V_{\bm{G}'-\bm{G}} 	\, c^n_{\bm{q-G}}=0 \nonumber
\end{align}
%
%
where $\bm{q}$ is the crystalline momentum defined in the BZ, and $n$ the band index that can vary between 1 and the total number of reciprocal lattice vectors $\bm{G}$ in Eq.~\eqref{periodic_pot}. Solving this set of $n$ linear equations results in the band structure $\mathcal{E}_n(\bm{q})$ for the periodic potential in Eq.~\eqref{periodic_pot} and the Bloch wave function given by
%
%
\begin{equation}\label{Bloch_states}
\psi_{n\bm{q}}(\bm{r})=\ee^{\ii \bm{q}\cdot\bm{r}} \sum_{\bm{G}} c^n_{\bm{q}-\bm{G}} \,\ee^{\ii \bm{G}\cdot\bm{r}}.
\end{equation}
%
%
In this expression, the term denoted by the summation defines the periodic part $u_{n\bm{q}}(\bm{r})$ of the Bloch wave function $\psi_{n\bm{q}}(\bm{r})$.
Finally, we can use the Bloch wave function in Eq.~\eqref{Bloch_states} to calculate the local density of states (LDOS)~\cite{monkhorst1976special}:
%
%
\begin{equation}
    \rho(\bm{r},\epsilon)=\sum_n \int_{\text{BZ}} \frac{d\bm{k}}{2\, \pi^2}\,\lvert \psi_{n \bm{k}}(\bm{r})\rvert^2\, \delta(\epsilon-\epsilon_{n \bm{k}}),
\end{equation}
%
%
which can be measured in the STM experiments~\cite{slot2017experimental}. We note on passing that integrating $\rho(\bm{r},\epsilon)$ over the unit cell results in the total DOS as in Eq.~\eqref{eq_DOS}.

%
%
\begin{table}[!t]
\centering
\begin{tabular}{|c | c | c | c | c | c|}
\hline
& $\epsilon/t$ & $\Delta/t$ & $t'/t$ & \#$_\text{CO}$ unit cell & $\square$ (nm$^2$)\\
\hline
CI & $-1.18182$ & $1.71499$ & $-0.87715$ &9 & 8 \\
CII & $-5.67708$ & $0.128125$ & $-0.428646$ &24 & 13\\ 
CIII & $-5.51934$ & $0.0431492$ & $-0.40663$ &15 & 18\\
\hline
\end{tabular}
\caption{Summary of the properties of the three configurations in Fig.~\ref{Fig_3}.}\label{Tab_1}
\end{table}
%
%
As stated before, we consider the three configurations of the dice lattice we presented in \textbf{Figure}~\ref{Fig_3}. These differ for the total number of CO molecules, 9 for the case of CI~\ref{Fig_3}(a), 24 for the case of CII~\ref{Fig_3}(b), and finally, 15 for the case of CIII~\ref{Fig_3}(c). A summary of the properties of the three configurations is given in \textbf{Table}~\ref{Tab_1}.  We present in \textbf{Figures}~\ref{Fig_4}(a),~\ref{Fig_4}(b), and~\ref{Fig_4}(c) the band structure for the three possible realizations of the dice lattice, CI, CII and CIII, respectively. Together with the band structure obtained by solving Eq.~\eqref{sys_equations}, we present the optimal fitting with the tight-binding model from Eq.~\eqref{BlochHam} at the $K$-point; it has to be noted that we include a homogeneous shift of the spectrum $\epsilon$. 

In \textbf{Figures}~\ref{Fig_4}(d),~\ref{Fig_4}(e), and~\ref{Fig_4}(f), we present the LDOS $\rho(\bm{r},E)$ at the center of the rim and hub sites for the three configurations, this quantity would be accessible via STM measurements. These results help us to understand the role played by the next-nearest-neighboring model ($t'\neq0$). As introduced before, the perfectly flat band case $t'=0$ can be characterized by a linear combination of CLSs around the hub sites~\cite{sutherland1986localization,Bergman_2008,graf2021designing,Zhang_2020}. Surprisingly, for $t'\neq0$, we can observe that the LDOS is still localized around the rim sites for energies corresponding to the $\mathcal{E}_\text{c}$ band | \textbf{Figure}~\ref{fig_2}(b) and ~\ref{fig_2}(d). The survival of  \emph{imperfect} CLSs is more evident from the plots of the LDOS in space \textbf{Figures}~\ref{Fig_4}(g),~\ref{Fig_4}(h), and~\ref{Fig_4}(i), these density plots for $\rho(\bm{r},E)$ are obtained for an energy corresponding crossing between the upper and the central band at the K point. Here, it is possible to recognize that the LDOS still has a shape resembling the CLSs around the hub sites, with values that are much larger on the rim sites compared to the hub sites~\cite{Bergman_2008}.

\subsection{Low-energy theory}\label{lwa}

In this section, we will present a low-energy, long-wavelength approximation for the Hamiltonian~\eqref{BlochHam} around the K/K$'$ points. In the limit of $t'=0$, the approximated Hamiltonian reads
%
%
\begin{equation}\label{ham_le}
\mathcal{H}_0 = v_\text{F}\, \bm{\Sigma}\cdot\bm{p},
\end{equation}
%
%
where, $v_\text{F}=3ta/2$ is the Fermi velocity, and
$\bm{p}=-\ii\hbar(\partial_x,\partial_y,0)$ is the momentum operator in the lattice $xy$-plane. We can observe that Eq.~\eqref{ham_le} bears a remarkable resemblance to the Hamiltonian governing the behavior of electrons in graphene.
In Eq.~\eqref{ham_le}, there is a significant difference as compared to the graphene one~\cite{CastroNeto_2009}. Here, the pseudo-spin vector $\mathbf{\Sigma}=(\Sigma_x,\Sigma_y,\Sigma_z)$ has a total spin $S=1$, which reflects that the dice lattice has three inequivalent sites per unit cell as compared to the two sites in graphene. These $3\times 3$ matrices can be expressed as a linear combination of Gell-Mann matrices~\cite{Gell-Mann_1962} $\lambda_i$ and read as follow: 
%
%
\begin{subequations}\label{matrices}
    \begin{align}
 \Sigma_x&=\frac{1}{\sqrt{2}}(\lambda_1+\lambda_6) =
 \frac{1}{\sqrt{2}}\begin{pmatrix}
 0 & 1 & 0 \\
 1 & 0 & 1 \\
 0 & 1 & 0
 \end{pmatrix},\\
 \Sigma_y&=\frac{1}{\sqrt{2}}(\lambda_2+\lambda_7) =\frac{1}{\sqrt{2}}\begin{pmatrix}
 0 & -\ii & 0 \\
 \ii & 0 & -\ii \\
 0 & \ii & 0
 \end{pmatrix}\!\!,
 \end{align}
 \begin{align}
 \Sigma_z&=(\lambda_3+\sqrt{3}\lambda_8)=\begin{pmatrix}
 1 & 0 & 0 \\
 0 & 0 & 0 \\
 0 & 0 & -1
 \end{pmatrix},
\end{align}
\end{subequations}
%
%
which satisfy angular momentum commutation relations $[S_i,S_j]=\ii \epsilon_{ijk}S_k$, with $\epsilon_{ijk}$ the Levi-Civita tensor~\cite{Bercioux_2009,Bercioux_2011,Bercioux_2017}. We can introduce a chiral symmetry operator defined as
%
%
\begin{align}
	\Gamma=\begin{pmatrix}
		1 & 0 & 0 \\
		0 & -1 & 0 \\
		0 & 0 & 1
	\end{pmatrix},
\end{align}
%
%
that is unitary $\Gamma^\dag\Gamma=\Gamma^2=\mathbb{I}_3$ where $\mathbb{I}_3$ is the identity matrix of dimension 3, and anticommutes with the Hamiltonian~\eqref{ham_le}: $\{\Gamma,\mathcal{H}_0\}=0$. This chiral symmetry operator governs the energy symmetry of the system, see Fig.~\ref{fig_1}(c)~\footnote{Here the chiral symmetry is simply anticommuting with the system Hamiltonian since the flat band is at zero energy. In other lattice systems with three sites in the unit cell, such as the kagome lattice, the flat band is at finite energy. Thus they require the introduction of a \emph{generalized} chiral symmetry~\cite{Ni_2018,Kempkes_2019,Herrera_2022}.}.
%
%
\begin{figure}[!t]
\centering
\includegraphics[width= .99\columnwidth]{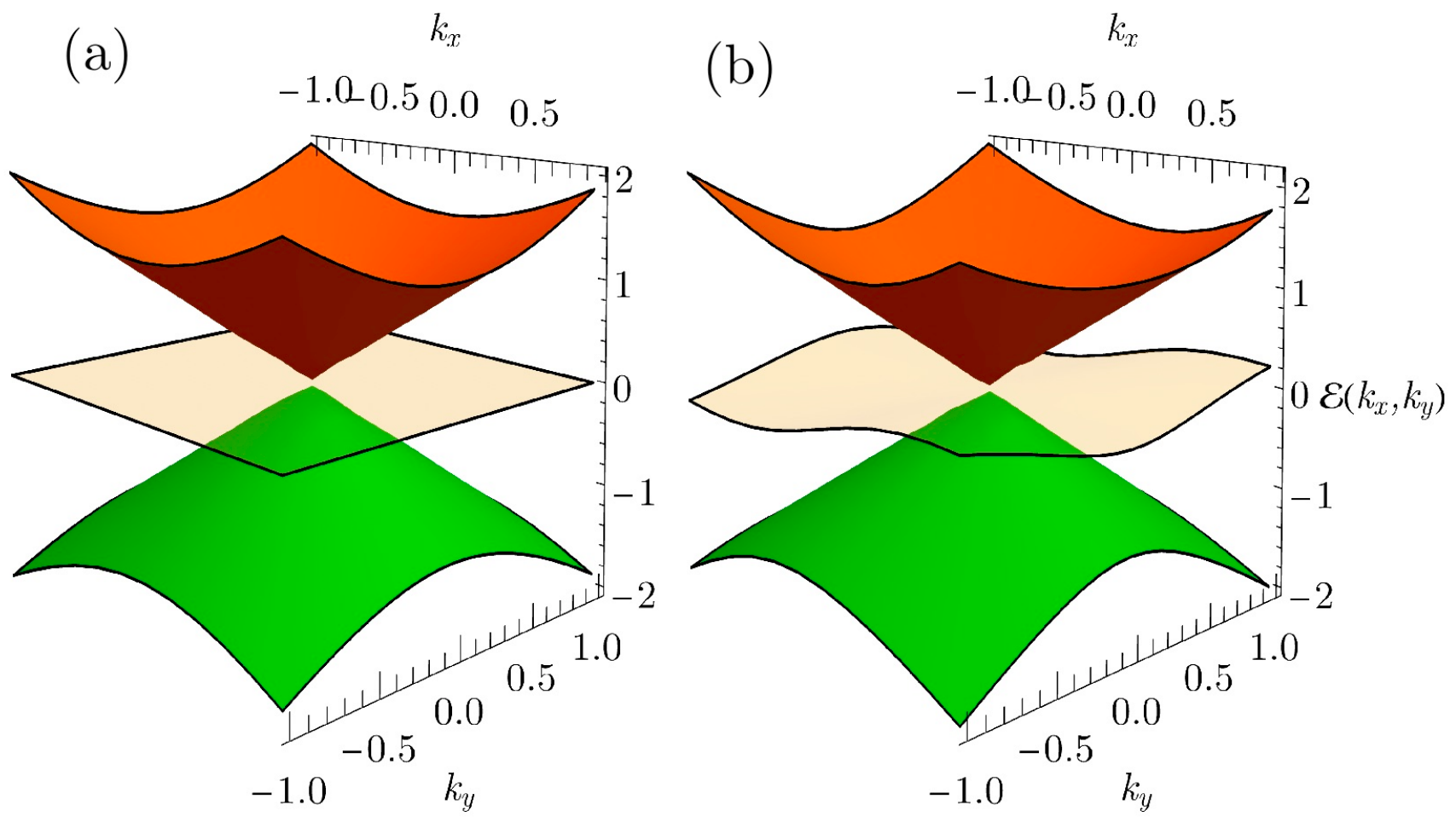}
\caption{\label{Fig_5} Energy spectrum as a function of the momenta in the long-wavelength approximation as in Eqs.~\eqref{ham_le} and~\eqref{ham_Lambda}. (a) case with $\Xi=0$, (b) case with $\Xi=0.1$.}
\end{figure}
%
%

We can complement the matrices~\eqref{matrices} with two additional Gell-Mann matrices, $\lambda_4$ and $\lambda_5$, so to describe the corresponding low-energy next-nearest-neighboring hopping term $t'$. This Hamiltonian term reads:
%
%
\begin{align}\label{ham_Lambda}
	\mathcal{H}_\Xi= \Xi (\lambda_4 p_x+\lambda_5 p_y),
\end{align}
%
%
where the coupling constant is $\Xi=3 t' a/2$. Importantly, this additional term breaks the chiral symmetry of the system: $\{\Gamma,\mathcal{H}_\Xi\}\neq0$ and thus the energy symmetry, see Fig.~\ref{fig_2}(c). For $\Xi=0$, we can write a simple expression for the energy spectrum:
%
%
\begin{align}\label{en_lwa}
	E^\text{lwa}_\alpha=\alpha v_\text{F} \sqrt{k_x^2+k_y^2},
\end{align}
%
%
with $\alpha\in\{0,\pm\}$, whereas, for $\Xi\neq0$, it is not possible to present a compact expression for the energy spectrum.
In \textbf{Figures}~\ref{Fig_5}(a) and~\ref{Fig_5}(b) we present the spectrum associated to Eqs.~\eqref{ham_le} and~\eqref{ham_Lambda} for the dice lattice when $\Xi$ is zero or finite, respectively. In \textbf{Figures}~\ref{Fig_5}(a), we can identify the three-fold degeneracy with the flat band in the $(0,0)$ point corresponding to the K point where we have performed the expansion of the Hamiltonian~\eqref{BlochHam}. In Figures~\ref{Fig_5}(b), for finite $\Xi$, we observe that the linear dispersion around the $(0,0)$ point is maintained but the flat band acquires a dispersion, and there is breaking of the energy symmetry.

\subsection{Magnetic field properties: tight-binding approach}\label{butterfly_tb}

%
%
\begin{figure}[t]
    \centering
    \includegraphics[width=\columnwidth]{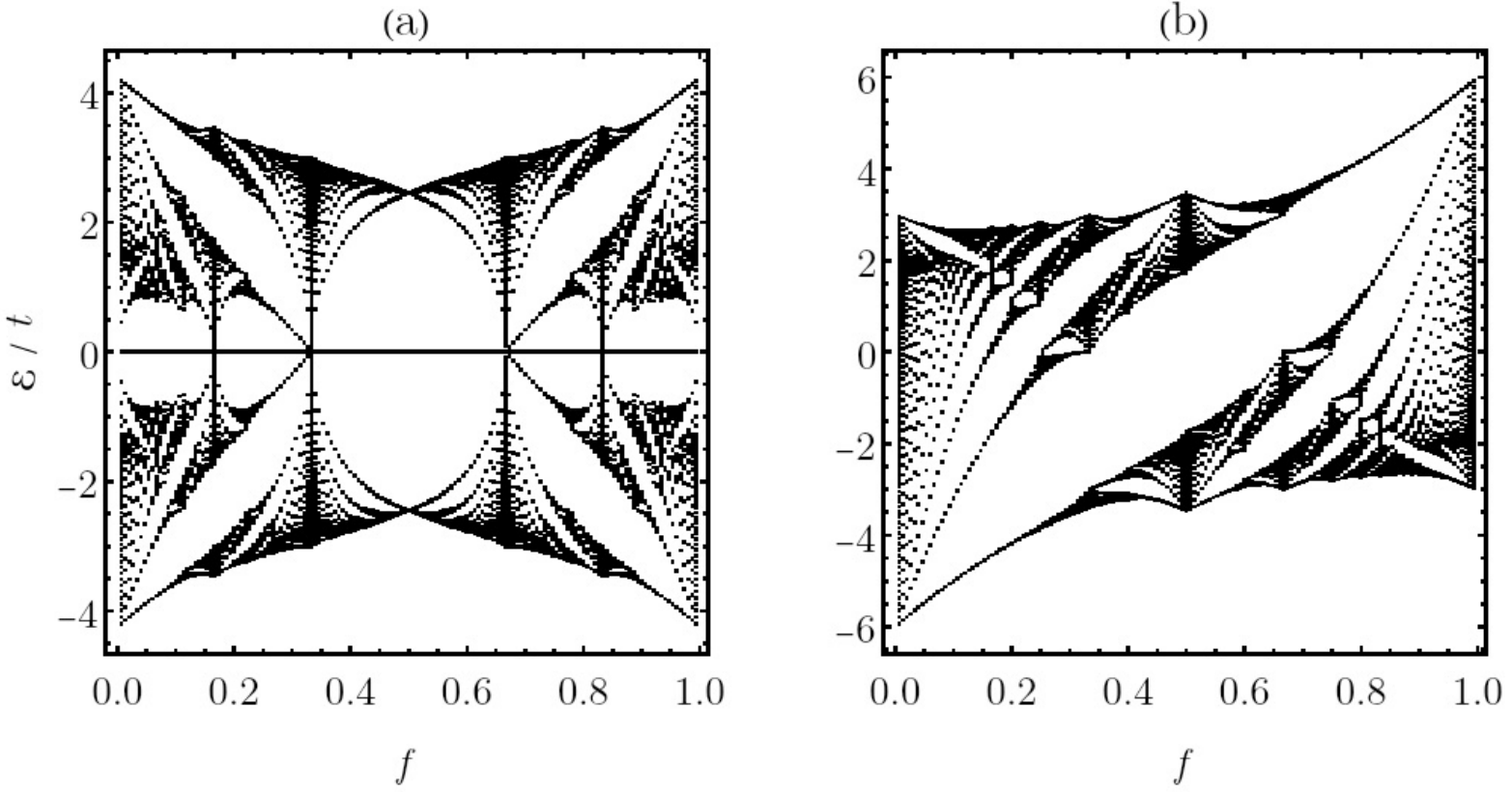}
    \caption{ \label{fig_6} Electronic energy spectrum as function of the magnetic flux  $f= B\, \sqrt{3}\, a^2/(2\, \phi_0)$ in natural units. (a)  Dice nearest neighbor case $t=1$, $t'=0$; (b) triangular lattice case $t=1$, $t'=-1$. }
\end{figure}
%
%
In the following, we show how to obtain the electronic energy spectrum in the presence of an external magnetic field in the tight-binding approximation. As customary, we perform the Peierls substitution in the hopping amplitude~\cite{peierls1933theorie}:
%
%
\begin{subequations}
\begin{align}
 t_{(\bm{R},\ell)\to (\bm{R}',\ell')} \to
 t_{(\bm{R},\ell)\to (\bm{R}',\ell')}\, \ee^{-\ii\, \frac{\hbar}{e}\, \Phi_{(\bm{R},\ell)\to (\bm{R}',\ell')}},
\end{align}
%
%
with
%
%
\begin{align}
 \Phi_{(\bm{R},\ell)\to (\bm{R}',\ell')} = \int_{\bm{R}_\ell}^{\bm{R}'_{\ell'}} d\bm{r}\cdot\bm{A}(\bm{r}),
\end{align}
\end{subequations}
%
%
where $\ell,\, \ell' =\{\text{R',H,R}\}$ and $\bm{R}_{\ell} = n_1\, \bm{a}_1 +n_2\, \bm{a}_2 + \bm{\delta}_\ell $ is the position of the $\ell$-th site in the unit cell, with $\bm{\delta}_\ell$ the displacement within the unit cell.

We choose the Landau gauge $\bm{A}=(0,B\, x,0)$. Then,  considering, for example, the $\hat{y}$-axis parallel to $\bm{a}_2$, the lattice periodicity is partially conserved along the $y$-direction.  
%
%
\begin{figure*}
\centering
\includegraphics[width=\textwidth]{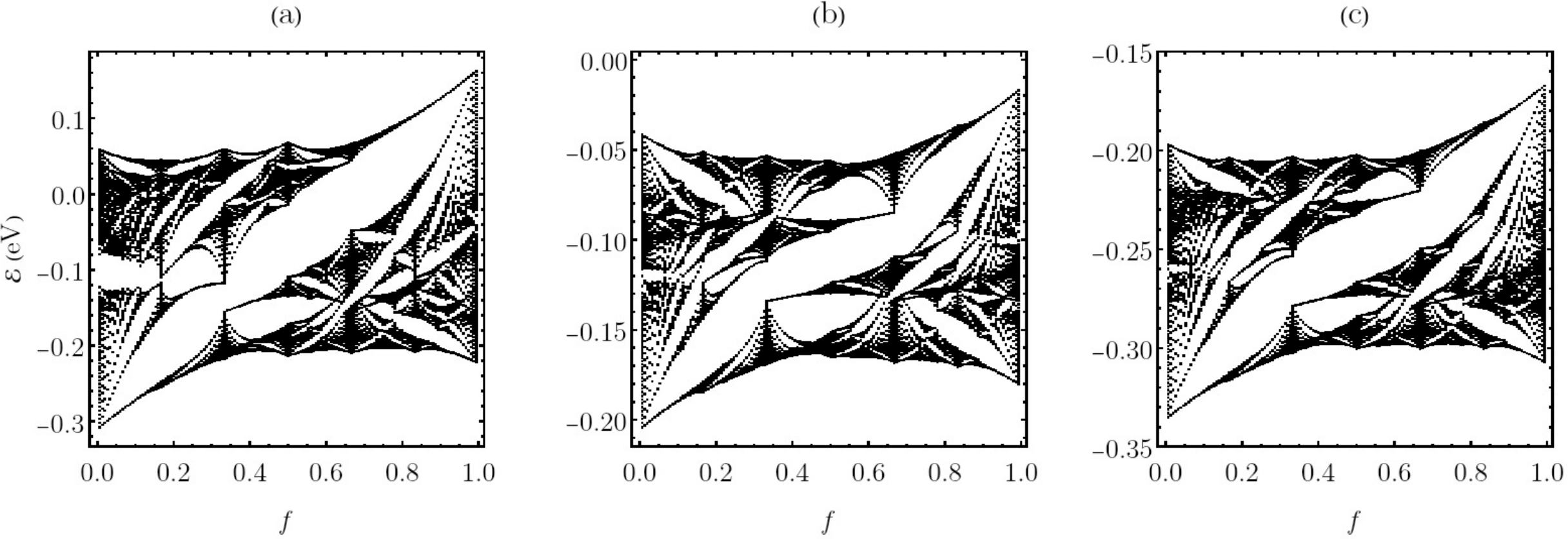}
\caption{\label{fig_7} Electronic energy spectrum as a function of the magnetic flux $f= B\, \sqrt{3}\, a^2/(2\, \phi_0)$ for the tight-binding model corresponding to the configuration (a) CI, (b) CII and (c) CIII, respectively.}
\end{figure*}
%
%
Within this choice of gauge, it is still possible to consider a change of basis 
%
%
\[\ket{n_1,n_2,l} = \frac{1}{\sqrt{N_2}}\sum_{k_2} \ee^{-\ii\, k_2\, n_2}\, \ket{n_1,k_2,l}\]
%
%
in which the tight-binding Hamiltonian reads
%
%
\begin{multline}
 \hat{H} =
 \sum_{k_2}\sum_{n_1}^{\text{PBC}} \Bigl[
 \bm{c}_{n_1,k_2}^\dagger \left( \mathcal{M} + \ee^{\ii\, k_2}\, \mathcal{P}_2 \right) \bm{c}_{n_1,k_2} +\\
 \bm{c}_{n_1,k_2}^\dagger \left( \mathcal{P}_1 + \ee^{-\ii\, k_2}\, \mathcal{P}_3 \right) \bm{c}_{n_1+1,k_2}
  + \text{h.c.}\Bigr],
\end{multline}
%
%
where $\bm{c}_{n_1,k_2}^\dagger=(\hat{c}^\dagger_{n_1,k_2,\text{R}},\, \hat{c}^\dagger_{n_1,k_2,\text{H}},\, \hat{c}^\dagger_{n_1,k_2,\text{R'}})$ are the creation operators,  the superscript in the $n_1$-summation refers to the periodic boundary conditions and
%
%
\begin{subequations}
\begin{align}
 \mathcal{M} &= \frac{t}{4}
 \begin{pmatrix}
  0 &  \ee^{-\frac{1}{2}\, \ii\, \pi  f (6 \, n_1-1)} & 0\\
  \ee^{\frac{1}{2}\, \ii\, \pi\,  f\, (6\, n_1-1)} & -\frac{\Delta}{t} &  \ee^{-\frac{1}{2}\, \ii\, \pi\,  f\, (6\, n_1+1)}\\
 0 & \ee^{\frac{1}{2}\, \ii\, \pi\,  f\, (6\, n_1+1)} &  0
 \end{pmatrix}, \label{Mmatrix}\\
 \mathcal{P}_1 & =
 \begin{pmatrix}
       0 & 0 & 0 \\
 t & 0 & 0 \\
 t'\, \ee^{\frac{3}{2}\, \ii\, \pi\,  f\, (2\, n_1+1)} & t & 0
\end{pmatrix},\\
\mathcal{P}_2 &=
\begin{pmatrix}
  0 & t\, \ee^{\frac{1}{2}\, \ii\, \pi\,  f\, (6\, n_1-1)} & t' \\
 0 & 0 & t\, \ee^{\frac{1}{2}\, \ii\, \pi\,  f\, (6\, n_1+1)} \\
 0 & 0 & 0
\end{pmatrix},\\
\mathcal{P}_3 &=
\begin{pmatrix}
  0 & 0 & 0 \\
 0 & 0 & 0 \\
 t'\, \ee^{-\frac{3}{2}\, \ii\, \pi\,  f\, (2\, n_1+1)} & 0 & 0
\end{pmatrix}
\end{align}
\end{subequations}
%
%
contain the magnetic hopping terms. We have omitted, for simplicity in Eq.~\eqref{Mmatrix}, a constant shift $\epsilon$ corresponding to the effective shift of the tight-binding model fitted over the three experimental proposals.

Typically, for the dice lattice, the magnetic flux is measured through an elementary rhombus (one-third of the unit cell surface) and normalized with respect to the $\phi_0= h/e$ magnetic flux quantum, that is  $f= B\, \sqrt{3}\, a^2/(2\, \phi_0)$. The calculation of the spectrum as a function of $f$ gives rise to the Hofstadter's butterflies~\cite{hofstadter1976energy} presented in \textbf{Figures}~\ref{fig_6} and~\ref{fig_7}.  In particular, \textbf{Figure}~\ref{fig_6}(a) represents the ideal
nearest neighbor $t=1$, $t'=0$ case~\cite{vidal1998aharonov}, whereas in \textbf{Figure}~\ref{fig_6}(b), we present the case of the triangular lattice corresponding to the  $t=1$, $t'=-t$ case~\cite{Du_2018}. Interestingly, for the case of the dice lattice at $f=1/2$,
all the states collapse in only three energy eigenvalues. This is a consequence of the Aharonov-Bohm caging mechanism~\cite{vidal1998aharonov}, for which the number of sites visited by
an initially localized wave packet is limited due to destructive interferences, resulting in an extreme localization phenomenon. 

As we have seen in \textbf{Section}~\ref{proposal}, due to the high mobility of the electrons of the Shockley state, we have finite values of the $t'$ parameter; consequently, the response in a magnetic field of our experimental proposal would substantially differ from the idea case in \textbf{Figure}~\ref{fig_6}(a). Considering the fitting parameters in \textbf{Table}~\ref{Tab_1},  we obtain the Hofstadter butterflies represented in \textbf{Figure}~\ref{fig_7}. We can clearly see how these fractal structures are an interpolation~\cite{barnsley1986fractal} of the pure Hofstadter butterfly for the case of the dice lattice in \textbf{Figure}~\ref{fig_6}(a) and the one for the triangular lattice in \textbf{Figure}~\ref{fig_6}(b).

\section{Conclusions and Outlook}\label{conclusions}

In this study, we explored the realization of a dice lattice on a Cu(111) surface decorated with CO molecules. Our experimental proposal demonstrates that the key spectral features of the dice lattice are maintained in some of the proposed implementations. We have verified this by fitting the electronic spectral properties using a next-nearest-neighbor tight-binding model. Some of the proposed implementations reveal a touching point and a quasi-flat band. We note on passing that one largest experiments performed with the electron quantum simulator for the realization of the Pensore tiling~\cite{Collins2017} included around 460 CO molecules. Extrapolating on this number would mean we could achieve around 20 unit cells for the dice lattice in the CII configuration. This number of unit cells is sufficient to already observe bulk properties in finite-size systems.

The finite bandwidth of the central quasi-flat band in our model is analogous to the Lieb lattice case, theoretically predicted~\cite{qiu2016designing} and experimentally implemented~\cite{slot2017experimental}. However, the actual value of the bandwidth of our proposal is of the order lower than $0.1$~eV in the two gapless configurations compared to the value of almost $0.2$~eV of the Lieb lattice. This proves that the experimental realization of the dice lattice can result in a quasi-flat band with a smaller bandwidth.
By studying the LDOS, we have shown a clear signature of the quasi-flat band in terms of a site-specific localization resembling CLSs; this spatial arrangement can be observed experimentally with a scanning tunneling microscope technique~\cite{slot2017experimental}.

As a final perspective for our results, we envision two important developments: as for the case of the honeycomb lattice, an opportune modulation of the lattice periodicity can result in a homogeneous shift of the Fermi energy~\cite{gomes2012designer}. This property can use used to create an effective $pn$-junction where the super-Klein tunneling~\cite{Urban_2011} could be observed. 
Secondly, to enhance the possibility of observing the effects of electron correlation in the flat band, we plan to investigate in future studies the implementation of the dice lattice in semiconducting electron simulator platforms based on InAs and InSb decorated with In or Cs adatoms~\cite{Pham_2022,Sierda_2023}. Finally, regarding the external magnetic field, an interesting outlook would be to study the strain-induced pseudomagnetic field in the case of next-nearest-neighbor term~\cite{PhysRevB.106.155417} since intensities of the order of 60 hundred Tesla have been achieved for the case the honeycomb lattice~\cite{gomes2012designer}.

%
%
\begin{figure}[!t]
    \centering
    \includegraphics[width=\columnwidth]{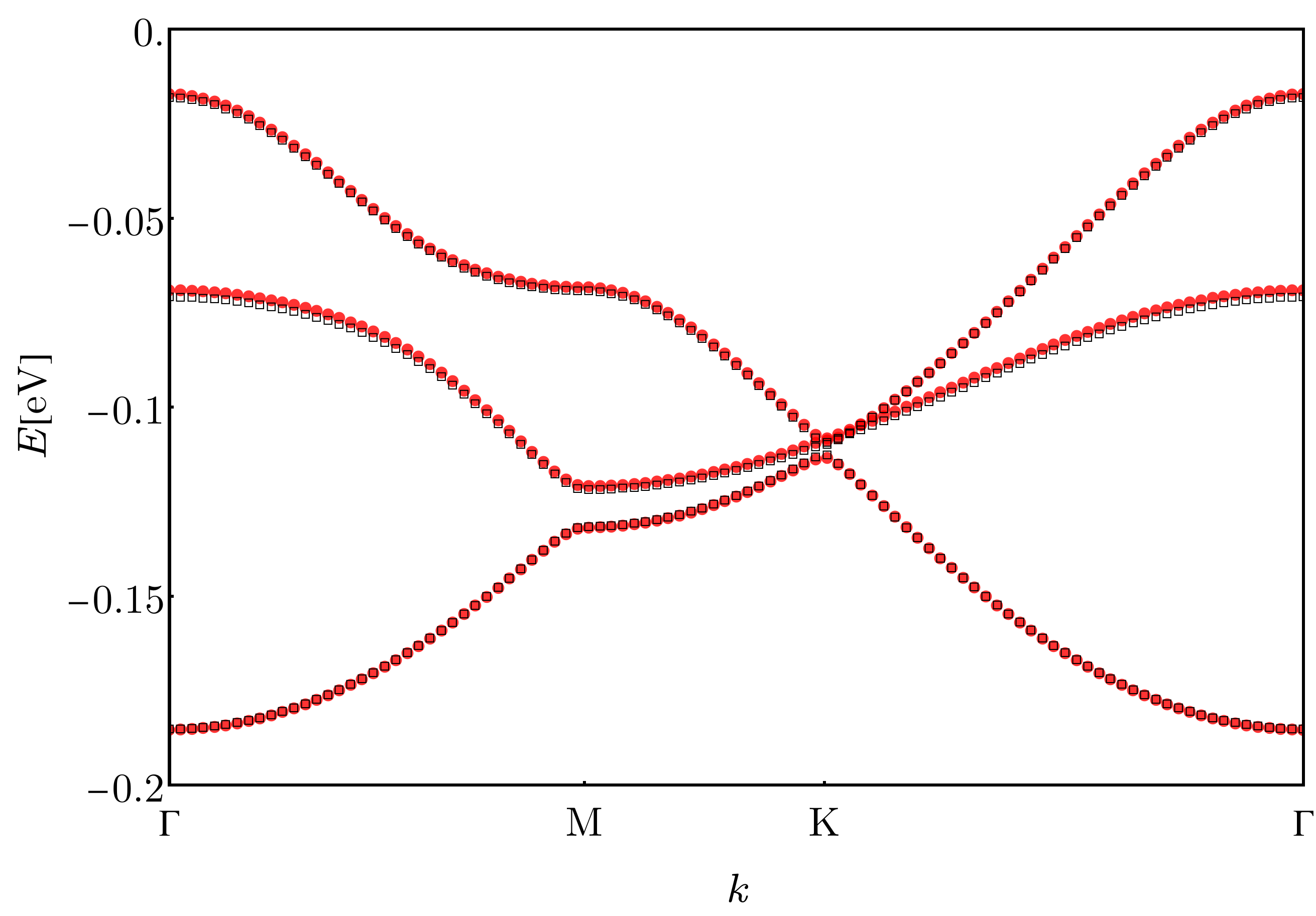}
    \caption{ \label{fig:Gaussian_pot} Electronic energy spectrum for the configuration with 24 CO molecules calculated with both cylindrical (empty square) and Gaussian (filled red-circles) potential.}
\end{figure}
%
%
\begin{acknowledgments} 
We acknowledge the useful discussion with Miguel Angel Cazalilla, Miguel Angel Jimenez Herrera, Rian Ligthart, and Ingmar Swart.
D.B. acknowledges the support from the Spanish MICINN-AEI through Project No.~PID2020-120614GB-I00~(ENACT), the Transnational Common Laboratory $Quantum-ChemPhys$,  the Department of Education of the Basque Government through
the project PIBA\_2023\_1\_0007 (STRAINER), and the financial support received from the IKUR Strategy under the collaboration agreement 
between the Ikerbasque Foundation and DIPC on behalf of the Department of Education of the 
Basque Government and the Gipuzkoa Provincial Council within the QUAN-000021-01 project.
\end{acknowledgments} 

\appendix

\section{Model and Parameters}\label{app_I}

Here, we discuss an additional shape for the muffin-thin potential and the dependence of the parameters for the case of the cylindrical one.

It turns out that the muffin-tin model with cylindrical potentials approximates very well the experimental results by choosing the parameters a potential barrier $V_0=0.9\, \text{eV}$ and a diameter $D=0.6\, \text{nm}$~\cite{gomes2012designer,slot2017experimental,Freeney_2022}. An alternative description of the CO molecules is a two-dimensional  isotropic Gaussian potential:
\begin{gather}
V_{\text{Gauss}}(\bm{r}-\bm{r}_0) = V_{0}^{(G)}\, \ee^{-\left(\frac{\bm{r}-\bm{r}_0}{\sqrt{2}\, \alpha}\right)^2},
\end{gather}
with $V_0^{(G)}=1.8\, \text{eV}$ and $\alpha=0.15\, \text{nm}$, such that 
%
%
\begin{align}
  \int_{\text{UC}} d\bm{r}\, V_{\text{Gauss}}^n(\bm{r}) \sim
  \int_{\textbf{R}^2} d\bm{r}\, V_{\text{Gauss}}^n(\bm{r})
  = \int_{\text{UC}} d\bm{r}\, V^n(\bm{r}),
\end{align}
%
%
where $n=1,2$ and $V$ is cylindrical potential. In particular, we can assume that the Gaussian vanishes fast enough within the unit cell and we require that the first and second momenta of the two potentials coincide. 

%
%
\begin{figure}[!t]
    \centering
    \includegraphics[width=\columnwidth]{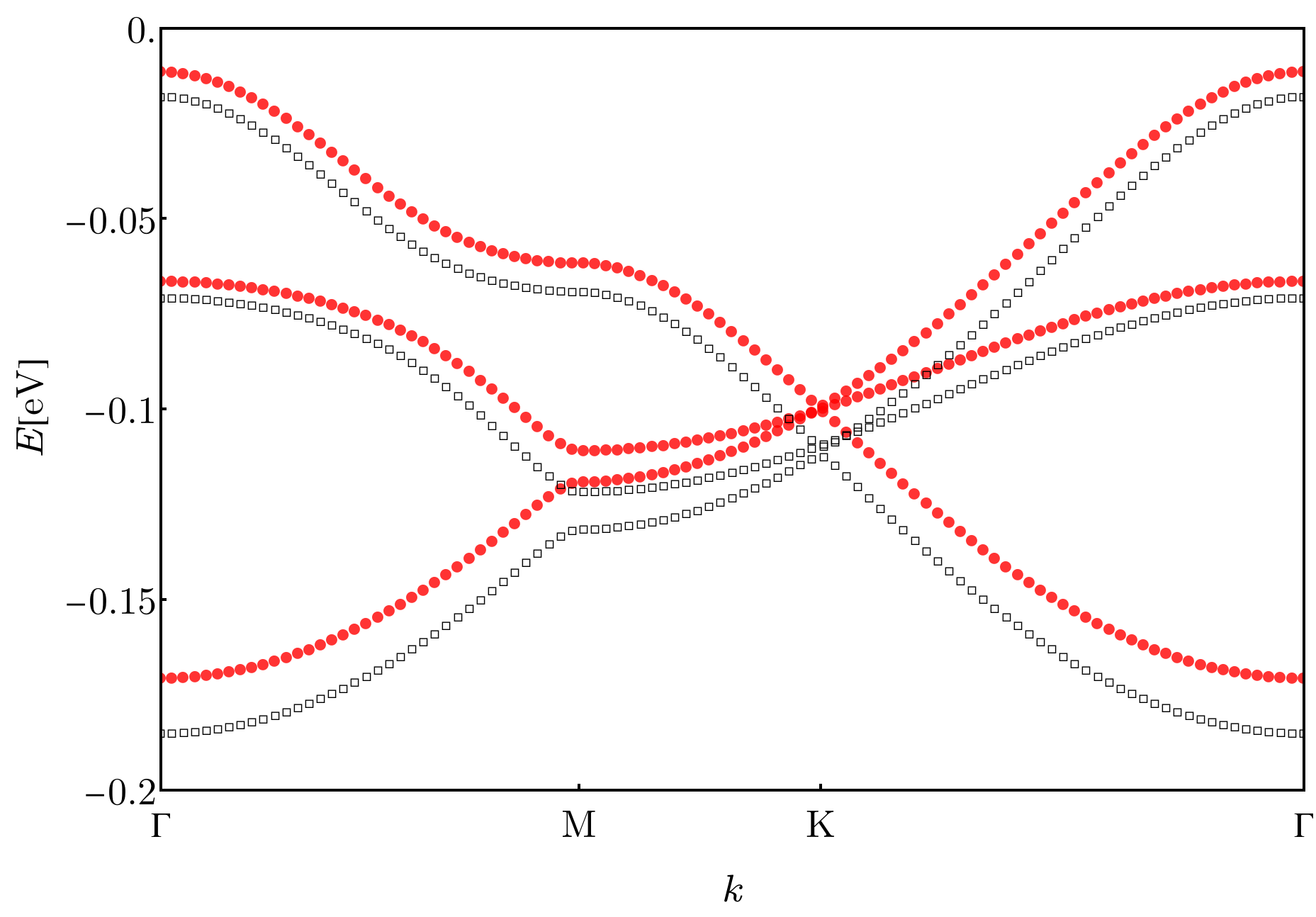}
    \caption{ \label{fig:highr_potential} Electronic energy spectrum for the configuration with 24 CO molecules calculated with both muffin-tin potential strength $V_0=0.9\, \text{eV}$ (empty square) and $V_0=1\, \text{eV}$ (filled red circles). The diameter potential is $D=0.6\, \text{nm}$ in both cases.
    }
\end{figure}
%
%

The Fourier components $V_{\bm{G}}$ are in the Gaussian case
%
%
\begin{subequations}
\begin{align}\label{Fourier_cyl}
V_{\bm{G}}& = \int_\text{UC} d\bm{r}\ V_{\text{Gauss}}(\bm{r}-\bm{r}_0)\text{e}^{-\text{i} \bm{G}\cdot \bm{r}} \nonumber\\ 
& = (2\pi \alpha^2 V_0) \ee^{-\lvert \bm{G}\rvert^2 \alpha^2/2}\,
\end{align}
\end{subequations}
and we found that the kind of potential does not affect significantly the nearly free electron model (in \textbf{Figure}~\ref{fig:Gaussian_pot} we report the case with 24 CO molecules).

Finally, by increasing the muffin-tin cylindrical potential strength $V_0$, we may expect higher values of the on-site energies and higher ratio $t/t'$ between the nearest and next-nearest-neighbor hopping terms, which is a measure of the central band flatness (note that $V_0$ is tunable only by changing the CO molecule with one with different repulsive potential~\cite{qiu2016designing}). This is shown in \textbf{Figure}~\ref{fig:highr_potential} for the 24 CO molecules case, where the potential has increased by about 10\%. This results in a lower gap $\Delta$ and in an increase of the $t/t'$ ratio by about 6\%.

\bibliography{bibliography}  

\end{document}